\documentclass[reqno,11pt]{article}

\usepackage[utf8]{inputenc}
\usepackage[T1]{fontenc}

\usepackage{amsmath}   
\usepackage{amssymb}  
\usepackage{amsthm}    
\usepackage{mathrsfs}  
\usepackage{mathtools} 
\usepackage{bm}

\usepackage{enumitem}  
\usepackage{graphicx} 
\usepackage{color}     
\usepackage[dvipsnames]{xcolor}
\usepackage{cite} 
\usepackage{authblk} 
\usepackage{hyperref}
\hypersetup{
	colorlinks=true,   
	linkcolor=blue,     
	citecolor=teal,    
	urlcolor=cyan,      
	pdftitle={}, 
	pdfauthor={}
}
\usepackage{cleveref}   

\usepackage[a4paper]{geometry}

\theoremstyle{plain} 
\newtheorem{theorem}{Theorem}
\newtheorem{lemma}{Lemma}
\newtheorem{corollary}{Corollary}
\newtheorem{proposition}{Proposition}

\theoremstyle{definition}
\newtheorem{definition}{Definition}
\newtheorem{example}{Example}

\Crefname{condition}{Condition}{Conditions}

\theoremstyle{remark}

\DeclareMathOperator{\diam}{diam}

\begin{document}

\title{Generalized Discrete Fractal Dimensions of Graph Families} 

\author[a]{Jiantao Liu \thanks{jiantaoliu@buaa.edu.cn}}
\author[b]{Cheng Zeng \thanks{Corresponding author: czeng@sdtbu.edu.cn}}
\author[a]{Yumei Xue \thanks{yxue@buaa.edu.cn}}
\author[a]{Qingcheng Zeng \thanks{qczeng@buaa.edu.cn}}

\affil[a]{{\small{School of Mathematical Science, Beihang University, Beijing  100191, P. R. China}}}
\affil[b]{{\small{ School of Mathematics and Information Science, Shandong Technology and Business University, 264003, P. R. China}}}

\date{}
\maketitle
\vspace{-2em}

\begin{abstract}
	We study generalized discrete fractal dimensions of sequences of finite connected graphs whose diameters tend to infinity.
	Using normalized shortest-path metrics and a general scale function, we define regularized energy dimensions for vertex probability measures and their optimized versions.
	We then establish a variational principle that identifies the optimized energy dimension with the supremum of ball-mass concentration dimensions.
	We also establish monotonicity in the moment order, stability under uniformly comparable measures, and upper bounds in terms of R\'enyi entropy growth and the upper mass dimension.
	Cluster estimates and explicit graph constructions show that local vertex concentration can lower the uniform energy dimension without changing the mass dimension.
	Finally, for uniformly Ahlfors-regular graph families, we obtain an integral criterion for energy boundedness and explicit formulas for the critical dimensions.
	These results clarify how global growth, local distance distributions, and the choice of scale function jointly determine discrete fractal dimensions.
	
	
\end{abstract}

\section{Introduction}
Fractal dimension quantifies how geometric complexity changes with scale. Classical fractal geometry provides complementary notions of dimension, with coverings, measures, and potentials describing different aspects of an irregular set \cite{falconer_fractal_2014, mandelbrot_how_1967}. For complex networks, Song, Havlin, and Makse \cite{song2005self} used renormalization by boxes to identify self-similar scaling, and subsequent work examined the origins of network fractality \cite{song2006origins} and the computation of box-covering exponents \cite{song2007calculate}. 
These works, together with the broader studies in \cite{bunimovich2024fractal, rosenberg_fractal_2020,wen2021fractal}, demonstrate that different notions of network dimension capture different aspects of network
structure. 
Understanding the relationships between these
dimensions is therefore an important problem.
%The approaches surveyed in \cite{bunimovich2024fractal, rosenberg_fractal_2020,wen2021fractal} show that a network dimension depends on both the geometric information being measured and the limiting procedure used to extract it. Establishing precise relationships between these quantities is therefore necessary for interpreting network scaling.

A finite graph, equipped with its shortest-path metric, has classical Hausdorff dimension zero. 
To obtain nontrivial dimensions, one can instead study a sequence of finite graphs
whose diameters tend to infinity.
%Nontrivial discrete dimensions can instead be assigned to a sequence of graphs whose diameters tend to infinity. 
Motivated by the potential-theoretic characterization of Hausdorff dimension associated with Frostman's work \cite{frostman1935potential}, Zeng and Xi \cite{zeng_hausdorff_2023} introduced an energy-based discrete Hausdorff dimension for families of networks. 
This approach has subsequently been applied to several recursively constructed network families, including flower networks and Hanoi graphs \cite{zeng2023hausdorff}, colored substitution networks \cite{zeng2025hausdorff}, and substitution tree networks \cite{zeng2026fractal}. This viewpoint makes the distribution of pairwise distances central to the dimension and provides a starting point for studying graph families without assuming that they arise from a fixed Euclidean fractal.

Extending this approach beyond specific recursive constructions raises a distinction between global growth and local concentration. 
Specifically, two graph families with comparable vertex counts and diameters may have quite different local vertex distributions. Such differences can substantially affect their energy dimensions.
Moreover, the choice of probability measure also plays a role. The uniform counting measure assigns equal weight to every vertex, whereas optimization over all probability measures allows us to investigate the best possible energy distribution on a graph family. These observations motivate a systematic study of the relationship between global growth, local mass concentration, and energy dimensions.
%The number of vertices and the diameter do not determine how many vertex pairs lie at short distances. 
%A second distinction concerns the measure: the uniform counting measure gives every vertex the same weight, whereas optimization over probability measures may place greater weight on less concentrated parts of a graph. Understanding these two effects requires estimates that connect pairwise interactions with the mass contained in small metric balls. Allowing scaling functions more general than the logarithm also makes it possible to investigate growth that is poorly described by a single power law.

A related development has taken place in classical fractal geometry, where generalized Hausdorff measures and multifractal dimensions have been studied using scale functions beyond the usual power functions \cite{rim2024some,rim2024A,selmi2025general,rim2024variational}. These theories provide a broader framework for describing scaling behavior and different distributions of mass. In the discrete setting, however, both the underlying metric space and the probability measure vary with the graph index. Consequently, extending these ideas to graph families requires estimates that remain uniform with respect to the graph index and the admissible scales.
%Generalized fractal measures provide a related line of development in the continuous setting. Achour and Selmi \cite{rim2024some} studied Hausdorff and packing constructions based on transformations of ball masses and radii, including comparisons between measures and pre-measures and questions of outer regularity. Achour, Li, Selmi, and Wang \cite{rim2024A} developed a general multifractal formalism and introduced relative box dimensions and general Rényi dimensions. These constructions illustrate how auxiliary functions accommodate scaling beyond the usual powers and how moment parameters distinguish different distributions of mass.
%Selmi \cite{selmi2025general} further investigated general multifractal dimensions of measures through their essential supports, regularity, singularity, and exact dimensionality. Achour, Li, and Selmi \cite{rim2024variational} established variational principles for general fractal dimensions and studied concentration dimensions, motivating our discrete formulation for graph families. In our setting, however, both the metric space and the measure vary with the graph index, and each normalized graph has a smallest positive distance. The required estimates must therefore control the graph index and the admissible radii simultaneously. We establish such estimates directly for graph sequences.

In this paper, we develop such a framework for sequences of finite connected graphs.
%We consider finite connected graphs with diameters tending to infinity and normalize shortest-path distances by the diameter. 
Let $\rho_n(x,y)=d_n(x,y) / D_n$ denote the normalized distance and let $\delta_n = D_n^{-1}$, where $D_n$ is the diameter. For a continuous strictly decreasing scale function $\psi$ that vanishes at one and tends to infinity at zero, we use the regularized kernel
$$ K_{s,n}^\psi(x,y):=\exp(s\psi(\max\{\rho_n(x,y),\delta_n\})).$$
The cutoff at \(\delta_n\) makes the diagonal contribution finite and allows us to optimize the corresponding energy over arbitrary vertex probability measures. When \(\psi(r)=\log(1/r)\), this framework recovers the usual power-scale energy and the associated discrete Hausdorff dimension. More general choices of \(\psi\) allow us to investigate scaling behavior that cannot necessarily be described by a single power law.
%The cutoff gives a finite diagonal contribution and permits optimization over atomic probability measures. Taking $\psi$ to be the logarithm of the reciprocal radius recovers the original discrete Hausdorff dimension for the uniform measure. We compare the energy dimensions with lower and upper mass dimensions, defined by the growth of the logarithm of the vertex count relative to $\psi$ evaluated at the cutoff \cite{rosenberg_fractal_2020}. These are growth exponents of graph families; their definition is distinct from the box-covering construction discussed above.

The main results of this paper can be summarized as follows.
First, we establish a discrete variational principle relating energy dimensions to the concentration of probability mass in metric balls. Under the assumption that $\psi(r)-\psi(2r)$ is bounded uniformly, we prove that the optimized energy dimension equals the supremum of the ball-mass concentration dimensions over all probability measure sequences. We also show that the energy dimension associated with the uniform counting measure does not exceed the optimized dimension. The proof relies on estimates connecting regularized potentials with ball masses and on a restriction argument that produces measures with controlled concentration.

Next, we compare energy dimensions with mass dimensions, which describe the growth of the number of vertices relative to the graph diameter. 
We prove that the optimized energy dimension is bounded above by the upper mass dimension. 
We further establish lower bounds for energy in terms of the size and diameter of concentrated vertex subsets. 
These estimates explain how local clustering can reduce the energy dimension even when the global mass growth remains unchanged. 
Explicit examples, including a path joined to a complete graph and a path with an attached binary heap tree, illustrate this phenomenon.

To examine mass distribution more closely, we introduce $q$-energy dimensions for $q>1$ through the moment of order $q-1$ of the regularized potential. These dimensions are nonincreasing in $q$ and invariant under changes of measure with uniformly bounded density ratios. Rényi entropy \cite{renyi1961measures} yields an upper bound through its growth relative to $\psi$ at the cutoff, and this entropy growth is itself bounded by the upper mass dimension. For the uniform measure, R\'enyi entropy equals the logarithm of the vertex count at every order. It therefore records global growth, while the energy also depends on distances between vertices. This comparison complements the covering-based information and generalized dimensions studied by Rosenberg \cite{rosenberg_maximal_2017, rosenberg_fractal_2020}.
In addition, we prove that optimization over probability measures removes the dependence on the moment order: the optimized \(q\)-energy dimension is the same for every \(q>1\).

Finally, we investigate the role of geometric regularity. For graph families satisfying a uniform Ahlfors regularity condition, we obtain two-sided estimates for the optimized \(q\)-energy and derive an integral criterion for its boundedness.
These results yield explicit formulas for the corresponding critical dimensions. In particular, for the power scale, the energy dimensions, concentration dimensions, mass dimensions, and R\'enyi entropy growth exponents all coincide with the Ahlfors regularity exponent. 
For more general scale functions, however, energy and mass dimensions need not agree, even under uniform Ahlfors regularity.

The paper is organized as follows. 
\Cref{sec: prenota} introduces normalized metrics, scale functions, and kernels. 
\Cref{sec:critial-dim} establishes the discrete variational principle.
\Cref{sec:mass-dim} compares energy dimensions with mass dimensions, while \Cref{sec:energy-bounds} develops energy estimates for concentrated vertex sets and provides explicit graph constructions. 
\Cref{sec:q-energy-dim} studies $q$-energy dimensions, R\'enyi entropy growth, and optimization over vertex measures. 
\Cref{sec:Ahlfors} investigates the relationship between energy dimensions and Ahlfors regularity.

\section{Preliminaries and notations}\label{sec: prenota}

For nonnegative sequences $\{a_n\}$ and $\{b_n\}$, we write
$a_n\lesssim b_n$ if there exists a constant $C>0$, independent of $n$, such that $a_n\le Cb_n$ for all sufficiently large $n$. We write $a_n\gtrsim b_n$ if $b_n\lesssim a_n$, and $a_n\asymp b_n$ if both $a_n\lesssim b_n$ and $a_n\gtrsim b_n$ hold. The implicit constants $C$ may depend on fixed parameters but not on $n$.

Let $\{G_n\}$ be a sequence of finite, connected, simple, unweighted graphs, where $G_n=(V_n,E_n)$. Write $d_n(x,y)$ for the shortest path length of $x$, $y\in V_n$ and assume  that $N_n:=|V_n|\geq 2$ is the order of $G_n$ and $D_n:=\operatorname{diam}G_n\to\infty$ is the diameter of $G_n$.

Set 
$$\delta_n=D_n^{-1}, ~\rho_n(x,y)=\frac{d_n(x,y)}{D_n},~B_n(x,r)=\{y\in V_n:\rho_n(x,y)\leq r\}.$$
Thus the normalized distance $\rho_n(x,y)$ lies in $[\delta_n,1]$ for $x\neq y$. Denote by $\mathcal P(V_n)$ the probability measures on $V_n$; here zero vertex weights are allowed.

The function $\psi:(0,1]\to[0,+\infty)$ is continuous and strictly decreasing, with $\psi(1)=0$ and $\psi(r)\to \infty$ as $r\downarrow 0$. In addition, let
\begin{equation}\label{eq:A_psi}
A_\psi:=\sup_{r\in(0,1/2]}\{\psi(r)-\psi(2r)\}<\infty.
\end{equation}
For $s\geq 0$, let $\phi_s(r):=\exp(-s\psi(r))$. Then
\begin{equation}\label{eq:phi_s}
\phi_s(\min\{2r,1\})\leq e^{sA_\psi}\phi_s(r), r\in(0,1].
\end{equation}
For $r\leq1/2$, \eqref{eq:phi_s} follows directly from \eqref{eq:A_psi}; for $r>1/2$, \eqref{eq:phi_s} holds since $\psi(r)\leq \psi(1/2)\leq A_\psi$.
\begin{example}
(Power scale) Set $\psi(r)=\log(1/r)$. We thus have $\phi_s(r)=r^s$ and $A_\psi=\log 2$.
\end{example}

For $\mu\in\mathcal{P}(V_n)$, define the distance energy 
$$\mathcal E_{s,n}^\psi(\mu):=\sum_{x\neq y}\mu(x)\mu(y)e^{s\psi(\rho_n(x,y))},$$
the regularized kernel
$$ K_{s,n}^\psi(x,y):=\exp(s\psi(\max\{\rho_n(x,y),\delta_n\})),$$
and the regularized energy
$$\mathcal J_{s,n}^\psi(\mu):=
\sum_{x,y\in V_n}\mu(x)\mu(y)K_{s,n}^\psi(x,y).$$
The regularization affects only the diagonal, that is,
\begin{equation}\label{eq:diag-decom}
	\mathcal J_{s,n}^\psi(\mu)=\mathcal E_{s,n}^\psi(\mu)
	+e^{s\psi(\delta_n)}\sum_x\mu(x)^2.
\end{equation}
In particular, $\mathcal J_{0,n}^\psi=1$, and $\mathcal J_{s,n}^\psi\ge1$ is
nondecreasing in $s$.

Let $\sigma_n$ be the uniform counting measure on $V_n$, i.e., $\sigma_n(x)=1/N_n$ for every $x\in V_n$.
\begin{proposition}\label{prop:uni}
	For every $n$ and $s\geq 0$, 
	\begin{equation}
	 \mathcal E_{s,n}^\psi(\sigma_n)\le\mathcal J_{s,n}^\psi(\sigma_n)
	\le2\mathcal E_{s,n}^\psi(\sigma_n).
	\end{equation}
\end{proposition}
\begin{proof}
	A connected graph $G_n$ has at least $N_n-1$ edges, so
	$$\mathcal E_{s,n}^\psi(\sigma_n)\geq
	\frac{2(N_n-1)}{N_n^2}e^{s\psi(\delta_n)}.$$
	The diagonal contribution therefore satisfies
	$$0\leq e^{s\psi(\delta_n)}\sum_x\sigma_n(x)^2=\frac{e^{s\psi(\delta_n)}}{N_n}
	\leq\frac{N_n}{2(N_n-1)}\mathcal E_{s,n}^\psi(\sigma_n)
	\leq\mathcal E_{s,n}^\psi(\sigma_n).$$
	Applying it to \eqref{eq:diag-decom}, we have the conclusion.
\end{proof}

In the power scale, the energy in the discrete Hausdorff dimension is 
\begin{equation}
	I_\alpha(G_n):=\mathcal E_{\alpha,n}^{\operatorname{pow}}(\sigma_n)
	=\frac1{N_n^2}\sum_{x\ne y}
	\left(\frac{D_n}{d_n(x,y)}\right)^\alpha.
\end{equation}
The discrete Hausdorff dimension $d_H$ is now the critical dimension of $I_\alpha(G_n)$, a special case of $\mathcal E_{s,n}^\psi(\mu)$.

\section{Critical dimension in general cases}\label{sec:critial-dim}
\begin{definition}\label{def:dim}
For a measure sequence $\mu_\bullet=(\mu_n)$ with $\mu_n\in\mathcal P(V_n)$, set 
\begin{align}
	d_u^\psi&:=\sup\{s\ge0:\liminf_{n\to\infty}\mathcal E_{s,n}^\psi(\sigma_n)<\infty\},\\
	d_{\mathrm{reg}}^\psi(\mu_\bullet)
	&:=\sup\{s\ge0:\liminf_{n\to\infty}\mathcal J_{s,n}^\psi(\mu_n)<\infty\}.
\end{align}
Set $a_n(s):=\min_{\mu\in\mathcal P(V_n)}\mathcal J_{s,n}^\psi(\mu)$. We can define another possibly larger dimension by
$$d_{\mathrm{ref}}^\psi:=\sup\{s\ge0:\liminf_{n\to\infty}a_n(s)<\infty\}.$$
\end{definition}
The notation ``ref'' refers to optimization of energy. The minimum $a_n(s)$ exists since the energy $\mathcal J_{s,n}^\psi(\mu)$ is a continuous function on the finite-dimensional compact probability simplex. Notice that dimensions $d_u^\psi$, $d_{\mathrm{reg}}^\psi(\mu_\bullet)$, $d_{\mathrm{ref}}^\psi$ take values in $[0,\infty]$. By convention, 
$$d_H=d_u^{\operatorname{pow}}=\sup\{\alpha\geq0:\liminf_{n\to\infty}	I_\alpha(G_n)<\infty\}.$$
\begin{definition}
Define  the maximum ball mass by
$$
Q_{n,\mu}(r):=\max_{x\in V_n}\mu(B_n(x,r)).
$$
The ball-mass concentration exponent is
\begin{equation}\label{eq:concen-dim}
	\begin{split}
		d_C^\psi(\mu_\bullet):=\sup\bigl\{t\ge0:\;&
		\text{there exist }n_k\uparrow\infty,\ C<\infty\text{ such that}\\
		&Q_{n_k,\mu_{n_k}}(r)\le C\phi_t(r)
		\quad\text{for all }k,\ \delta_{n_k}\le r\le1\bigr\}.
	\end{split}
\end{equation}
\end{definition}
Here  $C$ is independent of $k$ and $r$, but the constant $C$ and subsequence $\{n_k\}$ may
depend on $t$ and on the measure sequence $\mu_\bullet$. Remind that the smallest admissible radius of $Q_{n,\mu}(r)$ is $\delta_n$ and cannot tend to $0$ for a fixed $n$.

\begin{theorem}[Discrete variational principle]\label{thm:discrete dimensions}
For the dimensions defined in Definition \ref{def:dim}, we have
\begin{equation}\label{eq:variational}
	d_{\mathrm{ref}}^\psi
	=\sup_{\mu_\bullet}d_{\mathrm{reg}}^\psi(\mu_\bullet)
	=\sup_{\mu_\bullet}d_C^\psi(\mu_\bullet).
\end{equation}
Moreover,
\begin{equation}
	d_u^\psi=d_{\mathrm{reg}}^\psi(\sigma_\bullet)
	\le d_{\mathrm{ref}}^\psi.
\end{equation}
\end{theorem}
To prove the theorem above, we give some lemmas. We first show that $\mathcal{J}_{s,n}^\psi(\mu)$ is bounded if ball mass can be controlled.
\begin{lemma}\label{lem:ball-energy}
	Fix $n$ and $0<s<t$. If $\mu\in\mathcal P(V_n)$ satisfies
	\[
	Q_{n,\mu}(r)\le Ce^{-t\psi(r)},\qquad\delta_n\le r\le1,
	\]
	then
	\begin{equation}
		\mathcal J_{s,n}^\psi(\mu)\le1+\frac{Cs}{t-s}.
	\end{equation}
\end{lemma}
\begin{proof}
	Let $X,Y$ be independent vertices with distribution $\mu$, and put
	\[
	Z:=\psi(\max\{\rho_n(X,Y),\delta_n\}),\qquad L:=\psi(\delta_n).
	\]
	Then the decreasing property of $\psi$ implies $0\leq Z\leq L$. We also have  $\mathcal J_{s,n}^\psi(\mu)=\mathbb E\left( e^{sZ}\right)$.
	
	Using $$e^{sz}=1+\int_0^z se^{sy}\mathrm{d} y$$ and interchanging the nonnegative
	integral and expectation gives
	\begin{equation}\label{eq:mathcalJ}
		\mathcal J_{s,n}^\psi(\mu)
		=1+s\int_0^L e^{sy}\mathbb P(Z>y)\mathrm{d} y.
	\end{equation}
	
	If $L=0$, the claim is immediate. Otherwise, for $0\leq y<L$,
	continuity and strict monotonicity give a unique
	$r_y=\psi^{-1}(y)\in(\delta_n,1]$.
	The event $Z>y$ implies $\rho_n(X,Y)<r_y$, and hence
	$$
	\mathbb P(Z>y)
	\le\sum_x\mu(x)\mu(B_n(x,r_y))
	\le Q_{n,\mu}(r_y)\le Ce^{-ty}.
	$$
	We now derive that
	$$\mathcal J_{s,n}^\psi(\mu)
	\le1+Cs\int_0^L e^{-(t-s)y}\mathrm{d} y
	\le1+Cs\int_0^\infty e^{-(t-s)y}\mathrm{d} y
	=1+\frac{Cs}{t-s}.$$
\end{proof}
The following lemma shows that finite energy yields controlled ball mass.
\begin{lemma}\label{lem:energy-ball}
Fix $n$ and $s>0$. If $\mathcal J_{s,n}^\psi(\mu)\le M$, then there is $\nu\in\mathcal P(V_n)$ such that
\begin{equation}
Q_{n,\nu}(r)\le4Me^{sA_\psi}\phi_s(r),\qquad\delta_n\le r\le1.
\end{equation}
The measure $\nu$ is obtained by restricting $\mu$ to a set of $\mu$-mass at least $1/2$ and renormalizing.
\end{lemma}
\begin{proof}
$\mathcal J_{s,n}^\psi(\mu)\geq 1$ implies $M\geq 1$. Define
$$
U(x):=\sum_y\mu(y)K_{s,n}^\psi(x,y),\quad F:=\{x\in V_n:U(x)\le2M\}.
$$
Since $\sum_x\mu(x)U(x)=\mathcal J_{s,n}^\psi(\mu)\le M$, Markov's inequality
gives $\mu(F)\ge1/2$. Set $\nu(x)=\mu(x)\bm{1}_F(x)/\mu(F)$. 

Fix $z\in V_n$ and $\delta_n\le r\le1$, and write $B=B_n(z,r)$.
If $\nu(B)=0$, there is nothing to prove. Otherwise, choose
$x\in F\cap B$ with $\mu(x)>0$. For every $y\in B$, the triangle
inequality and the normalized diameter bound imply
$$
\max\{\rho_n(x,y),\delta_n\}\le R:=\min\{2r,1\}.
$$
Thus $K_{s,n}^\psi(x,y)\ge e^{s\psi(R)}=1/\phi_s(R)$, and
$$
2M\ge U(x)\ge\sum_{y\in B}\frac{\mu(y)}{\phi_s(R)}
=\frac{\mu(B)}{\phi_s(R)}.
$$
Using \eqref{eq:phi_s}, we conclude that 
$$
\nu(B)=\frac{\mu(F\cap B)}{\mu(F)}
\le2\mu(B)\le4M\phi_s(R)\le4Me^{sA_\psi}\phi_s(r).
$$
The estimate holds simultaneously for all centers and admissible radii.
\end{proof}
\begin{proof}[Proof of Theorem \ref{thm:discrete dimensions}]
For every $\mu_\bullet$ and $s\ge0$,
$a_n(s)\le\mathcal J_{s,n}^\psi(\mu_n)$. Hence every parameter admissible for
$d_{\mathrm{reg}}^\psi(\mu_\bullet)$ is admissible for
$d_{\mathrm{ref}}^\psi$, which gives
\[
\sup_{\mu_\bullet}d_{\mathrm{reg}}^\psi(\mu_\bullet)
\le d_{\mathrm{ref}}^\psi.
\]
Conversely, fix $s$ such that $\liminf_n a_n(s)<\infty$.
For each $n$, choose a minimizing measure $\mu_n^{(s)}$.
Then $\mathcal J_{s,n}^\psi(\mu_n^{(s)})=a_n(s)$, so
$d_{\mathrm{reg}}^\psi(\mu_\bullet^{(s)})\ge s$.
Taking the supremum over these $s$ proves the first equality in
\eqref{eq:variational}. 

Fix $\mu_\bullet$ and an admissible $t>0$ in
\eqref{eq:concen-dim}. There is a subsequence $n_k$ and a
uniform constant $C$ satisfying the ball-mass bound. For every $0<s<t$,
Lemma~\ref{lem:ball-energy} gives
\[
a_{n_k}(s)\le\mathcal J_{s,n_k}^\psi(\mu_{n_k})\le1+\frac{Cs}{t-s}.
\]
Therefore $d_{\mathrm{ref}}^\psi\ge s$. Letting $s\uparrow t$ and then
taking the supremum over $t$ and $\mu_\bullet$ yields
$d_{\mathrm{ref}}^\psi\geq\sup_{\mu_\bullet}d_C^\psi(\mu_\bullet)$.
The case $t=0$ is trivial.

Fix $s>0$ with $\liminf_n a_n(s)<\infty$.
Choose $M<\infty$ and $n_k\uparrow\infty$ so that $a_{n_k}(s)\le M$.
Apply Lemma \ref{lem:energy-ball} to minimizing measures at these
indices to obtain $\nu_{n_k}$ satisfying
\[
Q_{n_k,\nu_{n_k}}(r)\le4Me^{sA_\psi}\phi_s(r),
\qquad\delta_{n_k}\le r\le1.
\]
Define $\nu_n$ arbitrarily outside the subsequence, for example by
$\nu_n=\sigma_n$. This gives a full sequence with
$d_C^\psi(\nu_\bullet)\ge s$. Taking the supremum over admissible $s$
proves the reverse inequality $d_{\mathrm{ref}}^\psi\leq\sup_{\mu_\bullet}d_C^\psi(\mu_\bullet)$. If $d_{\mathrm{ref}}^\psi=0$, nonnegativity
settles this direction; if it is infinite, arbitrarily large admissible
parameters give an infinite supremum. The second equality in
\eqref{eq:variational} is proved now. 

The final statement follows from Proposition \ref{prop:uni}.
\end{proof}
\begin{proposition}
If
$$
M_s^\psi:=\int_0^1 e^{s\psi(x)}\mathrm d x<\infty,
$$
then there is a probability measure sequence $\lambda_\bullet$ for which
$\mathcal J_{s,n}^\psi(\lambda_n)\le3M_s^\psi$ for all $n$. Consequently,
\begin{equation}
	d_{\mathrm{ref}}^\psi\ge
	\sup\left\{s\ge0:\int_0^1e^{s\psi(x)}\mathrm d x<\infty\right\}.
\end{equation}
\end{proposition}
\begin{proof}
Fix a shortest path $v_0,\ldots,v_D$ realizing the diameter, where $D=D_n$. Hence $d_n(v_i,v_j)=|i-j|$.
Let $\lambda_n$ be uniform on these $D+1$ vertices and zero elsewhere. With $f_s(x)=e^{s\psi(x)}$, counting by distance gives
$$
 \mathcal J_{s,n}^\psi(\lambda_n)
=\frac{f_s(1/D)}{D+1}
+\frac2{(D+1)^2}\sum_{j=1}^D(D+1-j)f_s\left(\frac j D\right).
$$
Since $f_s$ is nonincreasing,
$$
\frac1D\sum_{j=1}^D f_s\left(\frac j D\right)\leq M_s^\psi,~
\frac1D f_s\left(\frac 1 D\right)\leq\int_0^{1/D}f_s(x)\mathrm d x\leq M_s^\psi.
$$
Hence $\mathcal J_{s,n}^\psi(\lambda_n)\leq 3M_s^\psi$. 

Taking the supremum over all $s\geq 0$ such that $M_s^\psi<\infty$ proves the second claim.
\end{proof}
\section{Comparison with mass dimension}\label{sec:mass-dim}
Let $L_n=\psi(\delta_n)\to\infty$. Define the generalized lower (upper) mass dimension by
\begin{equation}
	\underline d_B^\psi:=\liminf_{n\to\infty}\frac{\log N_n}{L_n},~
	\overline d_B^\psi:=\limsup_{n\to\infty}\frac{\log N_n}{L_n}.
\end{equation}
If they agree, their common value is denoted by $d_B^\psi$. This extends the growth convention in \cite{rosenberg_fractal_2020}. We claim that $\overline d_B^\psi$ is an upper bound for all dimensions we consider.
\begin{theorem}
\begin{equation}
	d_u^\psi\le d_{\mathrm{ref}}^\psi\le\overline d_B^\psi.
\end{equation}
\end{theorem}
\begin{proof}
The first inequality is from Theorem \ref{thm:discrete dimensions}. 

For any $\mu\in\mathcal P(V_n)$, the diagonal contribution and Cauchy--Schwarz give
\[
\mathcal J_{s,n}^\psi(\mu)\ge e^{sL_n}\sum_x\mu(x)^2
\ge\frac{e^{sL_n}}{N_n}.
\]
Consequently,
\[
a_n(s)\ge\exp\!\left[\left(s-\frac{\log N_n}{L_n}\right)L_n\right].
\]
Suppose $B:=\overline d_B^\psi<\infty$. Given $s>B$, choose
$0<\varepsilon<s-B$. For all sufficiently large $n$,
$\log N_n/L_n\le B+\varepsilon$, so
$a_n(s)\geq e^{(s-B-\varepsilon)L_n}\to\infty$.
Thus every $s>B$ is inadmissible and $d_{\mathrm{ref}}^\psi\leq B$.
If $B=\infty$, the inequality is trivial.
\end{proof}

\begin{figure}[h]
	\centering
	\includegraphics[width=0.9\textwidth]{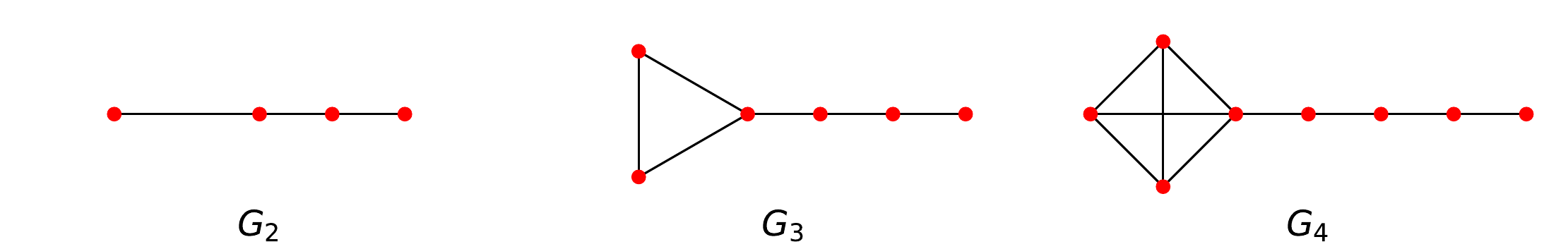}
	\caption{$\{G_n\}$ in Example \ref{ex:Kn-Pn}}
	\label{fig:Kn-Pn}
\end{figure}

As a corollary, we give the corresponding conclusion for traditional dimensions.
\begin{corollary}
\begin{equation}
	d_H\le d_{\mathrm{ref}}^{\operatorname{pow}}\le\overline d_B.
\end{equation}
\end{corollary}
\begin{proof}
It comes straightforwardly from
\begin{equation}\label{eq:cor}
\mathcal J_{s,n}^\psi(\sigma_n)\geq	 \mathcal E_{s,n}^\psi(\sigma_n)\geq\frac{2(N_n-1)}{N_n^2}e^{sL_n}\geq\frac{e^{sL_n}}{N_n}.
\end{equation}
\end{proof}
\begin{example}\label{ex:Kn-Pn}
We construct a family of networks $\{G_n\}$ as follows. 
For each $n \in \mathbb{N}$, the network $G_n$ is obtained by joining a complete graph $K_n$ and a path graph $P_n$, each consisting of $n$ vertices, by adding a single edge between one vertex of $K_n$ and one endpoint of $P_n$ (see Figure \ref{fig:Kn-Pn}).

It is straightforward to verify that the network $G_n$ has $N_n = 2n$ vertices and that its diameter satisfies $\diam(G_n) \asymp n$.
Hence, $d_B= 1$.
Furthermore, 
\begin{equation*}
	\begin{split}
	I_\alpha(G_n)
		\ge & \frac{1}{N_n^2} \sum_{\substack{u, v \in V(K_n) \\ u \ne v}}
		\left(\frac{\diam(G_n)}{d_n(u, v)} \right)^{\alpha} \\
		\asymp & \frac{1}{(2n)^2} \sum_{\substack{u, v \in V(K_n) \\ u \ne v}}
		n^{\alpha} \\
		\asymp & n^\alpha.
	\end{split}
\end{equation*}
Hence,
\begin{equation*}
	\liminf_{n \to \infty} 	I_\alpha(G_n) = \infty
	~ \text{for any}\ \alpha > 0.
\end{equation*}
Consequently, $d_H= 0$.
\end{example}
\begin{example}
For $k\ge1$, let $G_{2k-1}$ be the path with $k+1$ vertices, and let
$G_{2k}$ be the square grid on $\{0,\ldots,k\}^2$, with edges joining
vertices that differ by one in a single coordinate. Then $D_{2k-1}=k$ and $D_{2k}=2k$.

For every finite $k$ and $\alpha\in(0,2)$, we have 
$$
\sum_{j=1}^{2k}j^{1-\alpha}\leq C_\alpha k^{2-\alpha}.
$$
Then by a direct computation, 
\begin{align*}
	I_\alpha(G_{2k})
	&\leq\frac{(2k)^\alpha}{(k+1)^4}
	\sum_{x\in V_{2k}}\sum_{j=1}^{2k}4j\,j^{-\alpha}\\
	&=\frac{4(2k)^\alpha}{(k+1)^2}
	\sum_{j=1}^{2k}j^{1-\alpha}<\infty,
\end{align*}
which implies that $d_H\geq 2$.

Conversely, fix $\alpha>2$. The lower bound in \eqref{eq:cor} gives
$$
 I_\alpha(G_{2k-1})\ge\frac{k^\alpha}{k+1}\to\infty,~
I_\alpha(G_{2k})\ge\frac{(2k)^\alpha}{(k+1)^2}\to\infty.
$$
Both subsequences diverge, so $d_H\leq2$. 

We directly have 
\begin{align*}
\underline{d}_B&=\liminf_{n\to\infty}\frac{\log N_n}{\log D_n}=\lim\limits_{k\to\infty}\frac{\log N_{2k-1}}{ \log D_{2k-1}}=1,\\
\overline{d}_B&=\limsup_{n\to\infty}\frac{\log N_n}{\log D_n}=\lim\limits_{k\to\infty}\frac{\log N_{2k}}{\log D_{2k}}=2.
\end{align*}
This shows that $\underline{d}_B$ can be strictly smaller than $d_H$.
\end{example}

\section{Energy bounds for concentrated vertex sets}\label{sec:energy-bounds}
\begin{proposition}\label{prop:clusters}
	Let $S_n\subseteq V_n$, $m_n=|S_n|\ge2$, and suppose that $\operatorname{diam}(S_n):=\max_{x,y\in S_n} d_n(x,y)\leq r_n$, where $1\leq r_n\le D_n$.
	Put $p_n=m_n/N_n$ and $\varepsilon_n=r_n/D_n$.
	Then, for every $s\geq0$,
	\begin{equation*}
		\mathcal E_{s,n}^\psi(\sigma_n)\geq\frac12p_n^2e^{s\psi(\varepsilon_n)}.
	\end{equation*}
	Suppose that $S_{i,n}$, $1\le i\le M_n$, are pairwise disjoint sets satisfying
	the analogous conditions. With the corresponding notation, we set $m_{i,n}=|S_{i,n}|$, $\operatorname{diam}(S_{i,n})\leq r_{i,n}$, $p_{i,n}=m_{i,n}/N_{n}$ and $\varepsilon_{i,n}=r_{i,n}/D_n$. Then, for every $s\geq0$,
	\begin{equation*}
		\mathcal E_{s,n}^\psi(\sigma_n)\geq
		\frac12\sum_{i=1}^{M_n}p_{i,n}^2e^{s\psi(\varepsilon_{i,n})},
	\end{equation*}
\end{proposition}
\begin{proof}
	There are $m_n(m_n-1)$ ordered pairs in $S_n$, each with normalized distance at most
	$\varepsilon_n$ and hence kernel value at least $e^{s\psi(\varepsilon_n)}$. We thus derive
	\begin{equation*}
		\mathcal E_{s,n}^\psi(\sigma_n)
		\geq\frac{m_n(m_n-1)}{N_n^2}e^{s\psi(\varepsilon_n)}.
	\end{equation*}
	Using $1-1/m_n\ge1/2$, the first assertion follows.
	
	For disjoint sets $\{S_{i,n}\}$, the retained pair sets are disjoint, so their
	contributions may be added. Hence we obtain the second assertion.
\end{proof}
\begin{proposition}
If $\varepsilon_n\to0$, define
$$
\Lambda:=\limsup_{n\to\infty}
\frac{2\log(1/p_n)}{\psi(\varepsilon_n)}.
$$
Then
\begin{equation*}
	d_u^\psi\leq\Lambda.
\end{equation*}
\end{proposition}
\begin{proof}
If $\Lambda<\infty$ and $s>\Lambda$, choose
$0<\varepsilon<s-\Lambda$. For all sufficiently large $n$,
$$
\mathcal E_{s,n}^\psi(\sigma_n)
\geq\frac12e^{(s-\Lambda-\varepsilon)\psi(\varepsilon_n)}\to\infty,
$$
which implies that $d_u^\psi\le\Lambda$.

The case $\Lambda=\infty$ is trivial.
\end{proof}
\begin{corollary}\label{cor:d_H}
(1) Under Proposition \ref{prop:clusters}, if $p_n\geq c>0$ for all sufficiently large $n$ and $\varepsilon_n\to0$, then $d_u^\psi=0$. In particular, $d_H=0$.

(2) If $\varepsilon_n\to0$ and $p_n\asymp e^{-\theta\psi(\varepsilon_n)}$, then $d_u^\psi\leq2\theta$. In particular, if $\varepsilon_n\to0$ and $p_n\asymp\varepsilon_n^\theta$, then $d_H\leq2\theta$.
\end{corollary}
\begin{proof}
(1) For every $s>0$, Proposition \ref{prop:clusters} gives $\mathcal{E}_{s,n}^\psi(\sigma_n)\geq (c^2/2)e^{s\psi(\varepsilon_n)}\to\infty$. From the definition, we directly have $\mathcal{E}_{0,n}^\psi(\sigma_n)=1-1/N_n$. Hence $d_u^\psi=0$.

(2) Since $p_n\asymp e^{-\theta\psi(\varepsilon_n)}$, there exists a constant $c>0$ such that $p_n\geq ce^{-\theta\psi(\varepsilon_n)}$. Hence
\begin{equation*}
	\mathcal E_{s,n}^\psi(\sigma_n)\geq\frac12p_n^2e^{s\psi(\varepsilon_n)}\geq \frac{c^2}2e^{(s-2\theta)\psi(\varepsilon_n)}.
\end{equation*}
For $s>2\theta$, it is clear that $\mathcal E_{s,n}^\psi(\sigma_n)\to\infty$, which forces $d_u^\psi\leq2\theta$.

For $\alpha>2\theta$, $I_\alpha(G_n)\gtrsim\varepsilon_n^{2\theta-\alpha}\to\infty$, which implies that $d_H\leq2\theta$.
\end{proof}
\begin{corollary}
Suppose, for sufficiently large $n$, 
\begin{equation*}
N_n=D_n^{b+o(1)},~m_n=D_n^{a+o(1)},~r_n=D_n^{\gamma+o(1)},
\end{equation*}
where $b\geq 1$, $0\leq a\leq b$, and $0\leq \gamma<1$. Then
\begin{equation}\label{eq:prop3}
d_H\leq\min\left\{b,\frac{2(b-a)}{1-\gamma}\right\}.
\end{equation}
\end{corollary}
\begin{proof}
Using Proposition \ref{prop:clusters}, we obtain
$$
I_\alpha(G_n)\ge\frac12
D_n^{\alpha(1-\gamma)-2(b-a)+o(1)},
$$
which diverges when $\alpha>2(b-a)/(1-\gamma)$.

Referring to the proof of Proposition \ref{prop:uni}, we obtain
$$
 I_\alpha(G_n)\geq\frac{2(N_n-1)}{N_n^2}D_n^\alpha=D_n^{\alpha-b+o(1)},
$$
which diverges when $\alpha>b$. Combining the above two bounds yields \eqref{eq:prop3}.
\end{proof}
\begin{example}\label{ex:treefamily}
Fix $0<\eta\le1$. For $n\ge2$, take the path $P_n=(p_1,\ldots,p_n)$ and set
$m_n:=\max\{2,\lfloor n^\eta\rfloor\}$, $k_n:=\lfloor\log_2m_n\rfloor$.
Let $T_n$ be the binary heap tree on $\{1,\ldots,m_n\}$:
each $j\ge2$ is joined to $\lfloor j/2\rfloor$, with root $1$.
Its height is $k_n$. Adding the edge $\{p_1,1\}$ to connect $P_n$ and $T_n$, we obtain $G_n$. From $N_n=n+m_n\asymp n$, $D_n=n+k_n\asymp n$, we derive $d_B=1$.
	
The lower bound in \eqref{eq:cor} shows that 
$$I_\alpha(G_n)\ge\frac{2(n-1)D_n^\alpha}{N_n^2}\gtrsim n^{\alpha-1}.$$
Thus $d_H\leq 1$.
	
The tree cluster $T_n$ has diameter at most $2k_n\lesssim\log n$, so Proposition \ref{prop:clusters} gives
$$I_\alpha(G_n)\ge\frac{m_n(m_n-1)}{N_n^2}\left(\frac{D_n}{2k_n}\right)^\alpha\gtrsim\frac{n^{\alpha+2\eta-2}}{(\log n)^\alpha}.$$
For $\alpha>2-2\eta$, $I_\alpha(G_n)\to\infty$. This implies $d_H\leq 2-2\eta$.

To prove the reverse inequality, fix $\eta\in(0,1)$ and
$0<\alpha<\min\{1,2-2\eta\}$. The energy $I_\alpha(G_n)$ can be decomposed as
$$I_\alpha(G_n)=I_\alpha^{PP}(n)+I_\alpha^{TT}(n)+I_\alpha^{PT}(n),$$
where $I_\alpha^{PP}(n)$, $I_\alpha^{TT}(n)$, $I_\alpha^{PT}(n)$ represent the energy in $P_n$, $T_n$, and between $T_n$ and $P_n$, respectively. 

There are $2(n-j)$ ordered path pairs at distance $j$, so
\begin{align}
	I_\alpha^{PP}(n)
	&=\frac{2D_n^\alpha}{N_n^2}
	\sum_{j=1}^{n-1}(n-j)j^{-\alpha}\leq \frac{2nD_n^\alpha}{N_n^2}
	\sum_{j=1}^{n-1}j^{-\alpha}\\
	&\leq M_\alpha n^{\alpha-1}
	\sum_{j=1}^{n-1}j^{-\alpha}\leq M_\alpha C_\alpha.
\end{align}
Here integral comparison gives
$\sum_{j\le n}j^{-\alpha}\le C_\alpha n^{1-\alpha}$ and $M_\alpha$ is a constant depending only on $\alpha$.

All distinct tree vertices are at distance at least one. Hence
\begin{equation}
	I_\alpha^{TT}(n)\le\frac{D_n^\alpha}{N_n^2}m_n(m_n-1)
	\le C_\alpha n^{\alpha+2\eta-2}\to0.
\end{equation}

For a vertex $v\in T_n$ of depth $\operatorname{depth}(v)$, we have $d_n(p_i,v)=i+\operatorname{depth}(v)\geq i$. Thus 
\begin{align*}
	I_\alpha^{PT}(n)
	&=\frac{2D_n^\alpha}{N_n^2}
	\sum_{v\in T_n}\sum_{i=1}^n(i+\operatorname{depth}(v))^{-\alpha}\\
	&\leq\frac{2D_n^\alpha m_n}{N_n^2}\sum_{i=1}^n i^{-\alpha}
	\leq C_\alpha\frac{m_n}{n}\leq C_\alpha.
\end{align*}

Combining the above three estimates yields $I_\alpha(G_n)\le C_\alpha(1+n^{\alpha+2\eta-2})$. Every $\alpha<\min\{1,2-2\eta\}$ is therefore admissible. We now have $d_H=\min\{1,2-2\eta\}$, $\eta\in(0,1)$. 

When $\eta=1$, $I_\alpha(G_n)$ forces divergence for $\alpha>0$, while $I_0(G_n)$ is bounded. We thus have $d_H=0=\min\{1,2-2\eta\}$. 

In summary, $d_H=\min\{1,2-2\eta\}\leq d_B=1$ for graph family $\{G_n\}$. 

This also shows that for every \(h\in[0,1]\), there exists such a graph family with $d_H=h$ and $d_B=1$ (set $h=2-2\eta$).
\end{example}
\section{$q$-energy dimensions and R\'enyi entropy}\label{sec:q-energy-dim}
\begin{definition}
Fix $q>1$. For $\mu\in\mathcal P(V_n)$, define
the regularized potential and its $(q-1)$-st moment by
\begin{align}
	U_{s,n}^{\psi,\mu}(x)&:=\sum_{y\in V_n}\mu(y)K_{s,n}^\psi(x,y),\\
	\mathcal I_{q,s,n}^\psi(\mu)&:=
	\sum_{x\in V_n}\mu(x)\bigl(U_{s,n}^{\psi,\mu}(x)\bigr)^{q-1}.
\end{align}
The associated critical dimension of a measure sequence is
\begin{equation}\label{eq:q-dimension}
	d_{E,q}^\psi(\mu_\bullet):=
	\sup\{s\ge0:\liminf_{n\to\infty}\mathcal I_{q,s,n}^\psi(\mu_n)<\infty\}.
\end{equation}
\end{definition}
Because $K_{s,n}^\psi(x,y)\geq 1$ and $U_{s,n}^{\psi,\mu}(x)\geq 1$, we have $\mathcal I_{q,s,n}^\psi(\mu)\geq 1$. It is clear that $\mathcal I_{q,s,n}^\psi(\mu)$ is nondecreasing in $s$, and $\mathcal I_{q,0,n}^\psi(\mu)=1$. Thus $	d_{E,q}^\psi(\mu_\bullet)$ belongs to $[0,\infty]$.

At $q=2$, 
\begin{equation}
	\mathcal I_{2,s,n}^\psi(\mu)=\mathcal J_{s,n}^\psi(\mu),\qquad
	d_{E,2}^\psi(\mu_\bullet)=d_{\mathrm{reg}}^\psi(\mu_\bullet).
\end{equation}
Let $\sigma_\bullet=(\sigma_n)$ be the uniform counting measure on $\{G_n\}$. Proposition \ref{prop:uni} therefore implies
\begin{equation}\label{eq:dpowdh}
	d_{E,2}^\psi(\sigma_\bullet)=d_u^\psi,\qquad
	d_{E,2}^{\operatorname{pow}}(\sigma_\bullet)=d_H.
\end{equation}
The restriction $q>1$ is substantive. For $q=1$, $\mathcal I_{1,s,n}^\psi(\mu)=1$; for $q<1$, $U_{s,n}^{\psi,\mu}\geq1$ would imply
$0<\mathcal I_{q,s,n}^\psi\le1$. The resulting critical dimension would be infinite.
\begin{theorem}\label{thm:q-dim}
If $1<q_1<q_2$, then
\begin{equation}
	d_{E,q_1}^\psi(\mu_\bullet)\ge d_{E,q_2}^\psi(\mu_\bullet).
\end{equation}
\end{theorem}
\begin{proof}
Fix $n,s$, put $U(x)=U_{s,n}^{\psi,\mu}(x)$, and let
$a_i=q_i-1$. Since $0<a_1/a_2<1$, Jensen's inequality for the concave
function $z^{a_1/a_2}$ on the probability space $(V_n,\mu)$ gives
\begin{align}
	\mathcal I_{q_1,s,n}^\psi(\mu)
	&=\sum_x\mu(x)\bigl(U(x)^{a_2}\bigr)^{a_1/a_2}\\
	&\le\left(\sum_x\mu(x)U(x)^{a_2}\right)^{a_1/a_2}
	=\bigl(\mathcal I_{q_2,s,n}^\psi(\mu)\bigr)^{a_1/a_2}.
\end{align}
If $s$ is admissible for $q_2$, some subsequence has $q_2$-energy at most
a finite constant $M$. On the same subsequence, the $q_1$-energy is at most
$M^{a_1/a_2}$. Thus the admissible set for $q_2$ is contained in that for
$q_1$,as desired.
\end{proof}
By Theorem \ref{thm:q-dim} and \eqref{eq:dpowdh}, we straightforwardly have the following.
\begin{corollary}
\begin{equation}
	\begin{cases}
		d_{E,q}^\psi(\sigma_\bullet)\ge d_u^\psi,&1<q\le2,\\
		d_{E,q}^\psi(\sigma_\bullet)\le d_u^\psi,&q\ge2.
	\end{cases}
\end{equation}
\end{corollary}
The definitions below are R\'enyi entropy growth exponents.
\begin{definition}
For $q>1$, define
$$
	S_q(\mu):=\sum_x\mu(x)^q,~H_q(\mu):=-\frac{\log S_q(\mu)}{q-1}.
$$
With $L_n=\psi(\delta_n)\to\infty$, set
	\begin{equation}
		\underline B_q^\psi(\mu_\bullet):=\liminf_n\frac{H_q(\mu_n)}{L_n},~
		\overline B_q^\psi(\mu_\bullet):=\limsup_n\frac{H_q(\mu_n)}{L_n}.
	\end{equation}
\end{definition}
\begin{theorem}\label{thm:renyi-upper}
	For every $q>1$ and every measure sequence,
	\begin{equation}\label{eq:q-dim-mass}
		d_{E,q}^\psi(\mu_\bullet)
		\leq\overline B_q^\psi(\mu_\bullet)\leq\overline d_B^\psi.
	\end{equation}
\end{theorem}
\begin{proof}
Since $0\le\mu_n(x)\leq1$, $S_q(\mu_n)\le1$.
Jensen's inequality applied with equal weights to the $N_n$ numbers
$\mu_n(x)$ gives $S_q(\mu)\ge N_n^{1-q}$.
Hence $0\le H_q(\mu)\le\log N$, proving  $\overline B_q^\psi(\mu_\bullet)\leq\overline d_B^\psi$.

Notice that $U_{s,n}^{\psi,\mu_n}(x)\ge\mu_n(x)e^{sL_n}$. Hence
\begin{align*}
	\mathcal I_{q,s,n}^\psi(\mu_n)
	&\ge e^{s(q-1)L_n}S_q(\mu_n)\\
	&=\exp\!\left[(q-1)L_n
	\left(s-\frac{H_q(\mu_n)}{L_n}\right)\right].
\end{align*}
If $B:=\overline B_q^\psi(\mu_\bullet)<\infty$ and $s>B$, choose
$0<\varepsilon<s-B$. Eventually $H_q(\mu_n)/L_n\le B+\varepsilon$,
so $\mathcal I_{q,s,n}^\psi(\mu_n)\to\infty$. Thus $s$ is inadmissible. The case $B=\infty$ is trivial. We therefore conclude that $	d_{E,q}^\psi(\mu_\bullet)\leq\overline B_q^\psi(\mu_\bullet)$.
\end{proof}

For the uniform counting measure,
\begin{equation*}
	H_q(\sigma_n)=\log N_n,~\underline B_q^\psi(\sigma_\bullet)=\underline d_B^\psi,~
	\overline B_q^\psi(\sigma_\bullet)=\overline d_B^\psi.
\end{equation*}	
By analogy with Theorem \ref{thm:q-dim}, we obtain $H_{q_1}(\mu)\geq H_{q_2}(\mu)$ for $1<q_1<q_2$. Hence we have
\begin{theorem}
If $1<q_1<q_2$, then
\begin{equation*}
	\underline B_{q_1}^\psi(\mu_\bullet)\geq \underline B_{q_2}^\psi(\mu_\bullet),~\overline B_{q_1}^\psi(\mu_\bullet)\geq \overline B_{q_2}^\psi(\mu_\bullet).
\end{equation*}
\end{theorem}
\begin{theorem}\label{thm:ball-reg}
Suppose $d>0$ and there is a constant $C$ such that
for all sufficiently large $n$, all $x\in V_n$, and $\delta_n\le r\le1$,
\begin{equation}
	\mu_n(B_n(x,r))\leq Ce^{-d\psi(r)}.
\end{equation}
Then $d_{E,q}^\psi(\mu_\bullet)\geq d$ for every $q>1$.
If, in addition, $\overline B_q^\psi(\mu_\bullet)\leq d$, then
$d_{E,q}^\psi(\mu_\bullet)=d$ for that value of $q$.
\end{theorem}
\begin{proof}
Let $Y$ have distribution $\mu_n$ and put $Z=\psi(\max\{\rho_n(x,Y),\delta_n\})$. Similar to \eqref{eq:mathcalJ}, one has $$U_{s,n}^{\psi,\mu_n}(x)
=1+s\int_0^{L_n}e^{sy}\mathbb P(Z>y)\mathrm d y.$$
For $0\le y<L_n$, let $r_y=\psi^{-1}(y)\in(\delta_n,1]$.
The event $Z>y$ and decreasing property of $\psi$ imply $\rho_n(x,Y)<r_y$. Therefore
$\mathbb P(Z>y)\le Ce^{-dy}$, so
$$
 U_{s,n}^{\psi,\mu_n}(x)\leq 1+Cs\int_0^\infty e^{-(d-s)y}\mathrm d y=1+\frac{Cs}{d-s}.
$$
This formula is uniform in $x$ and $n$. Consequently,
$$
\mathcal I_{q,s,n}^\psi(\mu_n)\leq \left(1+\frac{Cs}{d-s}\right)^{q-1}.
$$
The admissibility of every $s\in(0,d)$ yields $d_{E,q}^\psi(\mu_\bullet)\geq d$. The equality $d_{E,q}^\psi(\mu_\bullet)=d$ follows from \eqref{eq:q-dim-mass}.
\end{proof}
\begin{corollary}
Let $B_{d_n}(x,R):=\{y\in V_n: d_n(x,y)\leq R\}$. Suppose that there exist constants $0<c\leq C$, $d>0$, and $n_0\in\mathbb N$ such that, for every
$n\ge n_0$, every $x\in V_n$, and every $1\leq R\leq D_n$,
$$
cR^d\le |B_{d_n}(x,R)|\le CR^d.
$$
Then $d_{E,q}^{\operatorname{pow}}(\sigma_\bullet)=d_B=d$ for every $q>1$.
\end{corollary}
\begin{proof}
Taking $R=D_n$ yields $cD_n^d\le N_n\le CD_n^d$, so $d_B=d$. 

For $\delta_n\leq r\leq1$, we take $R=rD_n$ to obtain
$$
\sigma_n(B_n(x,r))\leq\frac{C(rD_n)^d}{cD_n^d}
=\frac Cc r^d.
$$
Applying Theorem \ref{thm:ball-reg}, we have $d_{E,q}^{\operatorname{pow}}(\sigma_\bullet)=d$.
\end{proof}

We now show that these dimensions are stable under uniformly comparable measures.
\begin{theorem}
Suppose that $\mu_\bullet$ and $\nu_\bullet$ are two uniformly comparable measure sequences, i.e., 
$$c\nu_n(x)\leq \mu_n(x)\leq C\nu_n(x),$$
where $0<c\leq C$ are constants and the inequality holds for all admissible $n$, $x$. Then, 
\begin{equation}
	 d_{E,q}^\psi(\mu_\bullet)=d_{E,q}^\psi(\nu_\bullet),~
	\underline B_q^\psi(\mu_\bullet)=\underline B_q^\psi(\nu_\bullet),~
	\overline B_q^\psi(\mu_\bullet)=\overline B_q^\psi(\nu_\bullet).
\end{equation}
\end{theorem}
\begin{proof}
From the comparability of measures $\mu_n$ and $\nu_n$, we have 
\begin{equation*}
c^q\mathcal I_{q,s,n}^\psi(\nu_n)\leq\mathcal I_{q,s,n}^\psi(\mu_n)\leq C^q\mathcal I_{q,s,n}^\psi(\nu_n).
\end{equation*}
For each $s$, boundedness along a subsequence is therefore equivalent
for the two measure sequences $\mu_\bullet$ and $\nu_\bullet$, proving $d_{E,q}^\psi(\mu_\bullet)=d_{E,q}^\psi(\nu_\bullet)$.

Similarly, we obtain $c^qS_q(\nu_n)\le S_q(\mu_n)\le C^qS_q(\nu_n)$. So,
$$|H_q(\mu_n)-H_q(\nu_n)|
\leq\frac q{q-1}\max\{|\log c|,|\log C|\}.$$
After division by $L_n\to \infty$, we get 
\begin{equation*}
\frac{|H_q(\mu_n)-H_q(\nu_n)|}{L_n}\to0,
\end{equation*}
which implies that $\underline B_q^\psi(\mu_\bullet)=\underline B_q^\psi(\nu_\bullet)$, $\overline B_q^\psi(\mu_\bullet)=\overline B_q^\psi(\nu_\bullet)$.
\end{proof}
\begin{theorem}
Let $S_n\subseteq V_n$ satisfy $p_n=\mu_n(S_n)>0$ and normalized diameter $\operatorname{diam}(S_n)/D_n\leq \varepsilon_n\in[\delta_n,1]$. For $q>1$ and $s\geq0$, we have 
\begin{equation}\label{eq:q-cluster}
 \mathcal I_{q,s,n}^\psi(\mu_n)\geq p_n^q e^{s(q-1)\psi(\varepsilon_n)}.
\end{equation}
For pairwise disjoint clusters $S_{i,n}$, satisfying $p_{i,n}=\mu_n(S_{i,n})>0$ and $\operatorname{diam}(S_{i,n})/D_n\leq \varepsilon_{i,n}\in[\delta_n,1]$, we have 
\begin{equation}\label{eq:q-add}
	\mathcal I_{q,s,n}^\psi(\mu_n)\ge
	\sum_{i=1}^{M_n}p_{i,n}^q
	e^{s(q-1)\psi(\varepsilon_{i,n})}.
\end{equation}
\end{theorem}
\begin{proof}
Monotonicity of $\psi$ yields $U_{s,n}^{\psi,\mu_n}(x)\ge p_ne^{s\psi(\varepsilon_n)}$ for $x\in S_n$. 
Restricting the sum to $S_n$ therefore gives
\begin{equation*}
 \mathcal I_{q,s,n}^\psi(\mu_n)\geq\sum_{x\in S_n}\mu_n(x)U(x)^{q-1}\geq p_n\bigl(p_ne^{s\psi(\varepsilon_n)}\bigr)^{q-1}.
\end{equation*}
For pairwise disjoint clusters $S_{i,n}$, these lower bounds may be added, so we obtain \eqref{eq:q-add}.
\end{proof}
\begin{corollary}
If $\varepsilon_n\to0$ and
$$
\theta:=\limsup_{n\to\infty}\frac{\log(1/p_n)}{\psi(\varepsilon_n)},
$$
then
\begin{equation}
d_{E,q}^\psi(\mu_\bullet)\leq\frac q{q-1}\theta.
\end{equation}
In particular, $p_n\ge c>0$ forces $d_{E,q}^\psi(\mu_\bullet)=0$; $p_n\asymp e^{-\theta\psi(\varepsilon_n)}$ implies that $d_{E,q}^\psi(\mu_\bullet)\leq q\theta/(q-1)$.
\end{corollary}
\begin{proof}
The case $\theta=\infty$ is immediate. Otherwise, for $s>q\theta/(q-1)$, choose $\varepsilon>0$ such that $s(q-1)-q(\theta+\varepsilon)>0$. For all sufficiently large $n$, \eqref{eq:q-cluster} is bounded below by
\begin{equation*}
	e^{(s(q-1)-q(\theta+\varepsilon))\psi(\varepsilon_n)}\to\infty,
\end{equation*}
which implies that $d_{E,q}^\psi(\mu_\bullet)\leq q\theta/(q-1)$.

The special cases are analogous to Corollary \ref{cor:d_H}, and we omit the proof here.
\end{proof}
\begin{definition}
For $q>1$, put
\begin{equation}
A_{q,n}(s):=\min_{\mu\in\mathcal P(V_n)}\mathcal I_{q,s,n}^\psi(\mu),~d_{\mathrm{ref},q}^\psi:=\sup\{s\ge0:\liminf_n A_{q,n}(s)<\infty\}.
\end{equation}
\end{definition}
\begin{theorem}\label{thm:q-ref}
	For every $q>1$,
	\begin{equation}
		d_{\mathrm{ref},q}^\psi=\sup_{\mu_\bullet}d_{E,q}^\psi(\mu_\bullet)=d_{\mathrm{ref}}^\psi.
	\end{equation}
	If $A_\psi<\infty$, then their common value equals $\sup_{\mu_\bullet}d_C^\psi(\mu_\bullet)$.
\end{theorem}
\begin{proof}
Fix $n,s$. $\mathcal I_{q,s,n}^\psi(\mu)$ is continuous and thus its minimum exists.

Since $A_{q,n}(s)\leq\mathcal I_{q,s,n}^\psi(\mu_n)$, every parameter admissible
for a fixed sequence is admissible for the optimized dimension.
Conversely, fix $s$ admissible for the latter and select minimizing
measures at that same $s$ for every $n$. Their energy sequence equals
$A_{q,n}(s)$, so their dimension is at least $s$.
Taking the supremum over admissible $s$ proves $	d_{\mathrm{ref},q}^\psi=\sup_{\mu_\bullet}d_{E,q}^\psi(\mu_\bullet)$.

Fix $q>1$, put $a=q-1$, and suppose $\mathcal I_{q,s}^\psi(\mu)\leq M$.
Define
$$
F:=\{x:U_s^\mu(x)\le(2M)^{1/a}\},\qquad
\nu:=\frac{\mu|_F}{\mu(F)}.
$$
Markov's inequality gives
$\mu((U_s^\mu)^a>2M)\le1/2$, so $\mu(F)\ge1/2$.
Since $\nu\le2\mu$, for every $x\in\operatorname{supp}\nu$ we have
$$
U_s^\nu(x)\le2U_s^\mu(x)\le2(2M)^{1/a}=:T.
$$
For every other moment order $r>1$, it follows that
\begin{equation}
	\mathcal I_{r,s}^\psi(\nu)\le T^{r-1}.
\end{equation}

If $s$ is admissible for $d_{\mathrm{ref},q}^\psi$, choose minimizing
measures with $q$-energy at most $M$ along an infinite subsequence.
The preceding construction produces, on the same subsequence, measures
whose $r$-energy is at most the uniform constant $T^{r-1}$.
Thus $\liminf_n A_{r,n}(s)<\infty$.
Interchanging $q$ and $r$ shows that the admissible parameter sets coincide.
Taking $r=2$ yields
$d_{\mathrm{ref},q}^\psi=d_{\mathrm{ref}}^\psi$.

Under $A_\psi<\infty$, Theorem \ref{thm:discrete dimensions} supplies the equality
with $\sup_{\mu_\bullet} d_C^\psi(\mu_\bullet)$.
\end{proof}
\begin{example}
(1) For $\{G_n\}$ in Example \ref{ex:Kn-Pn}, we have $$d_{E,q}^{\operatorname{pow}}(\sigma_\bullet)=0<\underline{B}_q^{\operatorname{pow}}=\overline{B}_q^{\operatorname{pow}}=d_B=1.$$

For $p\in V(G_n\backslash K_n)$ and $x\in K_n$, let $\nu_n(p)=1/n$ and $\nu_n(x)=0$. For $s\in (0,1)$, we have
\begin{align*}
U_{s,n}^{\operatorname{pow},\nu_n}(p)\leq \frac{(n+1)^s}{n}\left(1+2\sum_{k=1}^{n-1}k^{-s}\right)
\leq \frac{(n+1)^s}{n}\left(1+\frac{2n^{1-s}}{1-s}\right)\leq C_{s,2},
\end{align*}
where $C_{s,2}$ is a constant depending on $s$. It follows that
$\sup_{n\ge2}\sup_{p\in V(P_n)}U_{s,n}^{\mathrm{pow},\nu_n}(p)\leq C_{s,2}$. Hence $\mathcal I_{q,s,n}^{\operatorname{pow}}(\nu_n)\leq C_{s,2}^{q-1}$, which means that all $s<1$ are admissible. 
So, $d_{E,q}^{\operatorname{pow}}(\nu_\bullet)\geq 1$. 

On the other hand, from Theorem \ref{thm:q-ref} and known dimension results,  we derive 
\begin{equation*}
d_{E,q}^{\operatorname{pow}}(\nu_\bullet)= d_{\operatorname{ref},q}^{\operatorname{pow}}=d_{\mathrm{ref}}^{\operatorname{pow}}=1.
\end{equation*}

(2) For the tree family in Example \ref{ex:treefamily}, we claim that the uniform counting measure satisfies 
\begin{equation}\label{eq:q-power}
	d_{E,q}^{\operatorname{pow}}(\sigma_\bullet)=\min\left\{1,\frac{q(1-\eta)}{q-1}\right\}
\end{equation}
for every $q>1$. 

Fix $q>1$ and put $a=q-1$. The diagonal and tree
contributions give, respectively,
$$\mathcal I_{q,s,n}^{\mathrm{pow}}(\sigma_n)
\gtrsim n^{a(s-1)},~\mathcal I_{q,s,n}^{\mathrm{pow}}(\sigma_n)
\gtrsim\frac{n^{as+q(\eta-1)}}{(\log n)^{as}}.
$$
Thus
$$
d_{E,q}^{\mathrm{pow}}(\sigma_\bullet)
\leq \min\left\{1,\frac{q(1-\eta)}{q-1}\right\}.
$$

For $0<\eta<1$, fix
$0<s<\min\{1,q(1-\eta)/(q-1)\}$ and set
\[
F_n=T_n\cup\{p_1,\ldots,p_{m_n}\}.
\]
The path contribution to the potential is uniformly bounded,
since
\[
\frac{D_n^s}{N_n}
\left(1+2\sum_{j=1}^{n}j^{-s}\right)\le C_s.
\]
The tree contribution is at most $D_n^s m_n/N_n$ everywhere.
For $p_i\notin F_n$, the bound $d_n(p_i,y)\ge i>m_n$
for $y\in T_n$ improves this to
\[
\frac{D_n^s m_n}{N_n i^s}
\le C(m_n/n)^{1-s}\le C.
\]
Since $\sigma_n(F_n)\lesssim m_n/n$, it follows that
\[
\mathcal I_{q,s,n}^{\mathrm{pow}}(\sigma_n)
\le C_{s,q,\eta}
\left[1+\frac{m_n}{n}
\left(\frac{D_n^s m_n}{N_n}\right)^a\right]
\le C'_{s,q,\eta}
\left(1+n^{as+q(\eta-1)}\right).
\]
This is uniformly bounded by the choice of $s$, proving the reverse inequality in \eqref{eq:q-power}.

By setting $\nu_n(p_i)=1/n$ and $\nu_n(x)=0$, $x\in T_n$, we can derive
\begin{equation*}
	d_{E,q}^{\operatorname{pow}}(\nu_\bullet)= d_{\operatorname{ref},q}^{\operatorname{pow}}=d_{\mathrm{ref}}^{\operatorname{pow}}=1.
\end{equation*}
\end{example}
\section{Dimensions under uniform Ahlfors regularity}\label{sec:Ahlfors}
We now give energy estimates under Ahlfors regularity. We say that $(G_n)$ is uniformly $d$-Ahlfors regular if
there exist $d>0$ and $0<c\le C$ such that,
for all sufficiently large $n$,
\[
cR^d\le |B_{n}(x,R)|\le CR^d,
\qquad x\in V_n,\quad 1\le R\le D_n.
\]
\begin{theorem}\label{thm:ahlfors-energy}
Suppose that $(G_n)$ is uniformly $d$-Ahlfors regular. Set
\[
	\Phi_{s,n}^{\psi}
	:=
	\delta_n^d e^{s\psi(\delta_n)}
	+d\int_{\delta_n}^{1}r^{d-1}e^{s\psi(r)}\,dr.
\]
Then, for every fixed $q>1$ and $s\ge0$,
	\[
	A_{q,n}(s)
	\asymp
	\mathcal I_{q,s,n}^{\psi}(\sigma_n)
	\asymp
	\bigl(\Phi_{s,n}^{\psi}\bigr)^{q-1},
	\]
where the comparison constants are independent of $n$.
	Moreover,
	\[
	\liminf_{n\to\infty}A_{q,n}(s)<\infty
	\quad\Longleftrightarrow\quad
	\int_0^1 r^{d-1}e^{s\psi(r)}\,dr<\infty.
	\]
The same equivalence holds with $A_{q,n}(s)$ replaced
	by $\mathcal I_{q,s,n}^{\psi}(\sigma_n)$.
\end{theorem}

\begin{proof}
	Throughout the proof, $n$ is sufficiently large.
	Taking $R=D_n$ gives $N_n\asymp D_n^d$, and hence
	\[
	\sigma_n(B_n(x,r))\asymp r^d,
	\qquad \delta_n\le r\le1.
	\]
	The same estimate holds for open balls when
	$\delta_n<r\le1$: use the closed ball of radius $r/2$
	if $r\ge2\delta_n$, and the singleton $\{x\}$ otherwise.
	
	Write $L_n=\psi(\delta_n)$. The layer-cake identity gives
	\[
	U_{s,n}^{\psi,\sigma_n}(x)
	=
	1+s\int_0^{L_n}
	e^{sy}\sigma_n\bigl(\{z:\rho_n(x,z)<\psi^{-1}(y)\}\bigr)
	\,dy.
	\]
	Consequently, uniformly in $x$,
	\[
	U_{s,n}^{\psi,\sigma_n}(x)
	\asymp
	1+s\int_0^{L_n}e^{sy}[\psi^{-1}(y)]^d\,dy
	=\Phi_{s,n}^{\psi}.
	\]
	The last equality follows by changing the order of
	integration. Averaging the $(q-1)$-st power proves
	\[
	\mathcal I_{q,s,n}^{\psi}(\sigma_n)
	\asymp(\Phi_{s,n}^{\psi})^{q-1}.
	\]
	
	We next obtain a lower bound for arbitrary measures.
	For every $\delta_n<r\le1$, there is a partition of $V_n$
	into at most $C_1r^{-d}$ sets of diameter less than $r$.
	Indeed, when $r\ge8\delta_n$, take a maximal
	$r/4$-separated set. Its $r/4$-balls cover $V_n$,
	whereas its $r/8$-balls are disjoint and each has
	$\sigma_n$-mass at least a constant multiple of $r^d$.
	Assigning each vertex to a covering ball gives the
	required partition. For $r<8\delta_n$, use singletons
	and $N_n\asymp\delta_n^{-d}$.
	
	If $\mu$ is any probability measure and $X,Y$ are
	independent with law $\mu$, this partition gives
	\[
	\mathbb P(\rho_n(X,Y)<r)
	\ge \sum_j\mu(E_j)^2
	\ge \frac1{\#\{E_j\}}
	\gtrsim r^d.
	\]
	Applying the layer-cake identity therefore yields
	\[
	\mathcal J_{s,n}^{\psi}(\mu)\gtrsim\Phi_{s,n}^{\psi},
	\qquad
	A_{2,n}(s)\gtrsim\Phi_{s,n}^{\psi}.
	\]
	
	For completeness, put $a=q-1$ and
	$M=\mathcal I_{q,s,n}^{\psi}(\mu)$. Define
	\[
	F=\{x:U_{s,n}^{\psi,\mu}(x)\le(2M)^{1/a}\},
	\qquad
	\nu=\frac{\mu|_F}{\mu(F)}.
	\]
	Markov's inequality gives $\mu(F)\ge1/2$. Thus
	$\nu\le2\mu$, and
	\[
	\mathcal J_{s,n}^{\psi}(\nu)
	\le2(2M)^{1/a}.
	\]
	It follows that
	\[
	A_{q,n}(s)\ge2^{-q}A_{2,n}(s)^{q-1}
	\gtrsim(\Phi_{s,n}^{\psi})^{q-1}.
	\]
	Together with
	$A_{q,n}(s)\le\mathcal I_{q,s,n}^{\psi}(\sigma_n)$,
	this proves the comparisons.
	
	Finally, let
	\[
	J_s=\int_0^1r^{d-1}e^{s\psi(r)}\,dr.
	\]
	If $J_s<\infty$, monotonicity of $\psi$ implies
	\[
	\delta_n^d e^{s\psi(\delta_n)}
	\le d\int_0^{\delta_n}r^{d-1}e^{s\psi(r)}\,dr,
	\]
	so $\Phi_{s,n}^{\psi}\le dJ_s$.
	If $J_s=\infty$, the integral over $[\delta_n,1]$
	tends to infinity. The asserted equivalences follow.
\end{proof}

\begin{theorem}\label{thm:oscillating-scale}
Under the assumptions of Theorem~\ref{thm:ahlfors-energy}, suppose that
	\[
	\kappa_+
	:=
	\limsup_{r\downarrow0}
	\frac{\psi(r)}{\log(1/r)}
	\in(0,\infty).
	\]
	Then, for every $q>1$,
	\[
	d_u^\psi
	=d_{\mathrm{reg}}^\psi(\sigma_\bullet)
	=d_{E,q}^\psi(\sigma_\bullet)
	=d_{\mathrm{ref},q}^\psi
	=d_{\mathrm{ref}}^\psi
	=d_C^\psi(\sigma_\bullet)
	=\frac d{\kappa_+}.
	\]
	In contrast, the mass dimensions and entropy growth
	exponents satisfy
	\[
	\underline B_q^\psi(\sigma_\bullet)
	=\underline d_B^\psi
	=d\liminf_{n\to\infty}
	\frac{\log D_n}{\psi(1/D_n)},
	\]
	\[
	\overline B_q^\psi(\sigma_\bullet)
	=\overline d_B^\psi
	=d\limsup_{n\to\infty}
	\frac{\log D_n}{\psi(1/D_n)}.
	\]
\end{theorem}

\begin{proof}
	Put $g(t)=\psi(e^{-t})$. By
	Theorem~\ref{thm:ahlfors-energy}, admissibility of $s$
	is equivalent to
	\[
	\int_0^\infty e^{sg(t)-dt}\,dt<\infty.
	\]
	If $s<d/\kappa_+$, choose $\lambda>\kappa_+$ with
	$s\lambda<d$. Eventually $g(t)\le\lambda t$,
	so the integral converges.
	
	If $s>d/\kappa_+$, choose
	$d/s<\lambda<\kappa_+$. There exist $t_j\to\infty$
	such that $g(t_j)\ge\lambda t_j$. Since $g$ is increasing,
	\[
	\int_{t_j}^{t_j+1}e^{sg(t)-dt}\,dt
	\ge e^{(s\lambda-d)t_j}
	\int_0^1e^{-du}\,du\longrightarrow\infty.
	\]
	Thus the critical exponent is $d/\kappa_+$.
	This proves the energy-dimension assertions; the equality
	for $d_u^\psi$ follows from the comparison of uniform
	off-diagonal and regularized energies.
	
	For the concentration dimension, the ball-mass estimate
	$\sigma_n(B_n(x,r))\asymp r^d$ shows that every
	$t<d/\kappa_+$ is admissible. Conversely, if $t$ is
	admissible along a subsequence, then, for every fixed
	$0<r\le1$, passage to sufficiently large indices gives
	\[
	cr^d\le C_t e^{-t\psi(r)}.
	\]
	Taking logarithms and then the upper limit as $r\downarrow0$
	yields $t\kappa_+\le d$.
	
	Finally, $N_n\asymp D_n^d$ and
	$H_q(\sigma_n)=\log N_n$ give the remaining formulas.
\end{proof}

\begin{example}
	Consider three classical self-similar network families: skeleton networks of the Sierpi\'nski gasket, the Sierpi\'nski hexagon, and the Lindstr\"om snowflake. 
	
	Let \(m\) denote the number of self-similar copies in each construction, and let \(\lambda\) be the linear scaling factor (i.e., the reciprocal of the contraction ratio). 
	See \Cref{fig:SierpinskiGasket,fig:SierpinskiHexagon,fig:Snowflake} for the parameters of these three network families.
	The discrete Hausdorff dimensions of these network families are given by \cite{zeng_hausdorff_2023}
	\[
	d = \frac{\log m}{\log\lambda}.
	\]
	Furthermore, these network families are uniformly \(d\)-Ahlfors regular.
	
	\begin{figure}[h]
		\centering
		\includegraphics[width=0.75\textwidth]{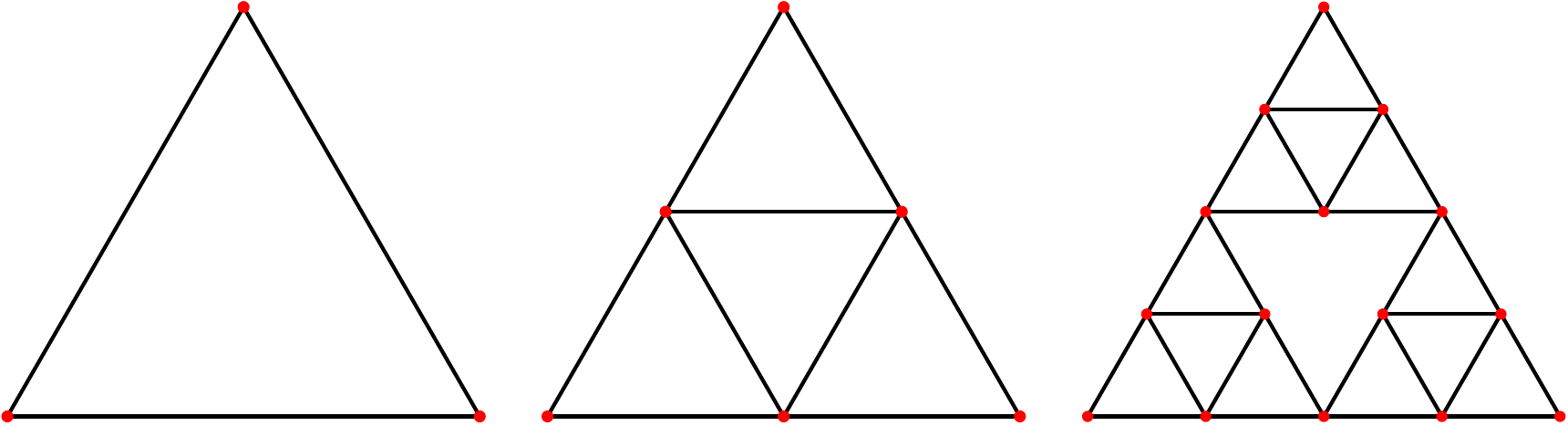}
		\caption{Skeleton networks of Sierpi\'nski gasket ($m = 3, \lambda = 2$)}
		\label{fig:SierpinskiGasket}
	\end{figure}
	\begin{figure}[h]
		\centering
		\includegraphics[width=0.75\textwidth]{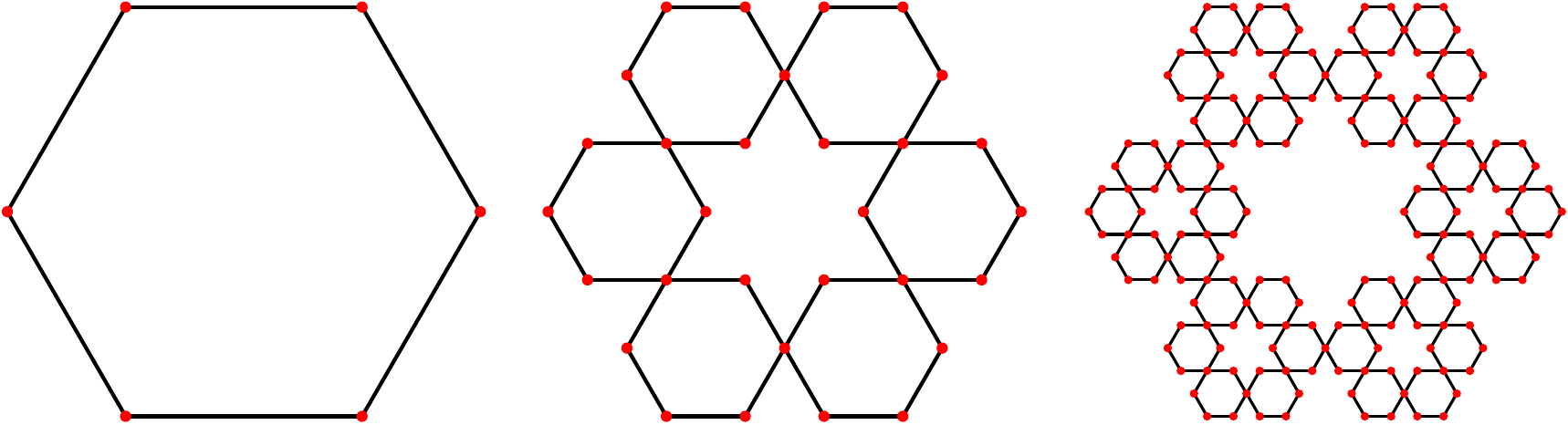}
		\caption{Skeleton networks of Sierpi\'nski hexagon ($m = 6, \lambda = 3$)}
		\label{fig:SierpinskiHexagon}
	\end{figure}
	\begin{figure}[h]
		\centering
		\includegraphics[width=0.75\textwidth]{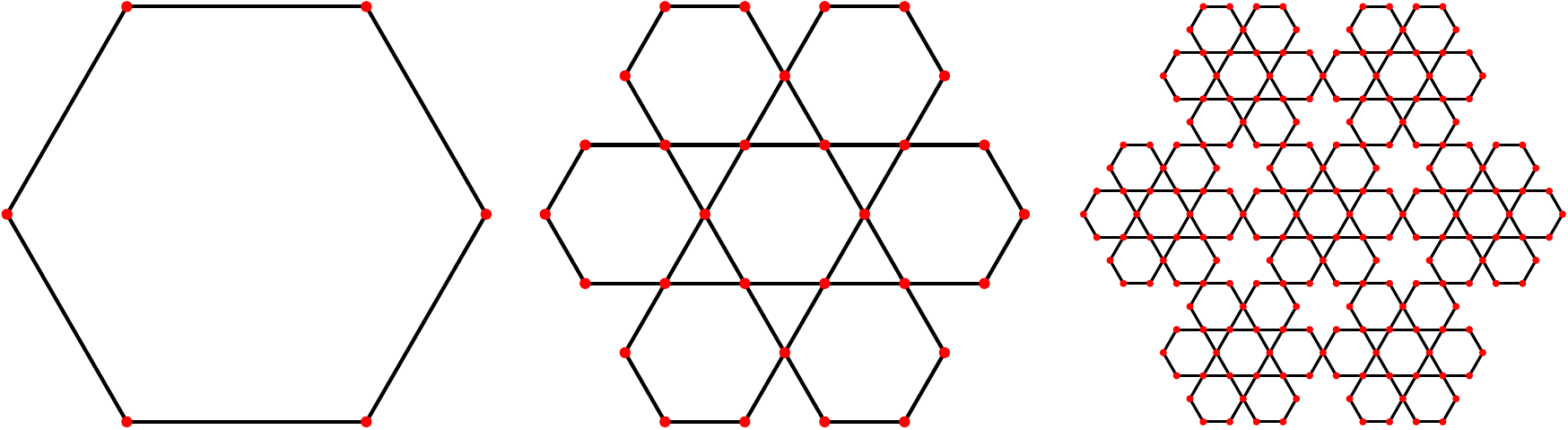}
		\caption{Skeleton networks of Lindstr\"om snowflake ($m = 7, \lambda = 3$)}
		\label{fig:Snowflake}
	\end{figure}
	
	Let \(\sigma_n\) be the uniform counting measure
	on \(V_n\), and take
	\[
	\psi(r)=\log\frac{1}{r}.
	\]
	Then \(A_\psi=\log 2<\infty\) and
	\[
	\kappa_+
	=
	\limsup_{r\downarrow0}
	\frac{\psi(r)}{\log(1/r)}
	=1.
	\]
	By \Cref{thm:oscillating-scale}, for every \(q>1\),
	\[
	d_u^\psi = d_{\mathrm{reg}}^\psi(\sigma_\bullet) = d_{E,q}^\psi(\sigma_\bullet) = d_{\mathrm{ref},q}^\psi = d_{\mathrm{ref}}^\psi = d_C^\psi(\sigma_\bullet) = d.
	\]
   Since \(|V_n|\asymp D_n^d\)
	and \(H_q(\sigma_n)=\log |V_n|\), we also have
	\[
	d_B^\psi = \overline d_B^\psi = B_q^\psi(\sigma_\bullet) = \overline B_q^\psi(\sigma_\bullet) = d.
	\]
	Therefore, for each of these three classical network families, all the generalized dimensions considered above coincide with its discrete Hausdorff dimension.
\end{example}
\begin{example}
	Let $T_0=1$ and $T_j=2^{2^j}$ for $j\ge1$.
	Define a continuous piecewise linear function $g$ by
	$g(t)=t$ on $[0,1]$, and let its slope on
	$(T_{j-1},T_j)$ be $3$ for odd $j$ and $1$ for even $j$.
	Set
	\[
	\psi(r)=g(\log(1/r)).
	\]
	Since $T_{j-1}/T_j\to0$,
	\[
	\frac{g(T_{2n})}{T_{2n}}\to1,
	\qquad
	\frac{g(T_{2n-1})}{T_{2n-1}}\to3.
	\]
	Also $1\le g(t)/t\le3$, so $\kappa_+=3$.
	The slope bound gives
	\[
	\psi(r)-\psi(2r)\le3\log2,
	\qquad 0<r\le1/2.
	\]
	Thus $\psi$ also satisfies the bounded-increment condition.
	
	Let $G_n$ be the path with diameter
	$D_n=\lfloor e^{T_{2n}}\rfloor$.
	This family is uniformly $1$-Ahlfors regular.
	Since $g$ is $3$-Lipschitz,
	\[
	\frac{\psi(1/D_n)}{\log D_n}\longrightarrow1.
	\]
	Therefore Theorem~\ref{thm:oscillating-scale} gives
	\[
	d_u^\psi
	=d_{E,q}^\psi(\sigma_\bullet)
	=d_{\mathrm{ref},q}^\psi
	=d_{\mathrm{ref}}^\psi
	=d_C^\psi(\sigma_\bullet)
	=\frac13,
	\]
	whereas
	\[
	\underline B_q^\psi(\sigma_\bullet)
	=\overline B_q^\psi(\sigma_\bullet)
	=\underline d_B^\psi
	=\overline d_B^\psi=1.
	\]
	Hence even uniform Ahlfors regularity and bounded scale
	increments do not force equality between generalized
	energy and mass dimensions.
\end{example}
\section{Conclusions}
We developed generalized discrete fractal dimensions of graph
families through regularized energies and their $q$-energy
extensions. Under the bounded-increment condition on $\psi$,
the discrete variational principle gives
$$
d_{\mathrm{ref},q}^{\psi}=\sup_{\mu_\bullet}d_{E,q}^{\psi}(\mu_\bullet)=d_{\mathrm{ref}}^{\psi}=\sup_{\mu_\bullet}d_C^{\psi}(\mu_\bullet),~ q>1.
$$

We established monotonicity in $q$, invariance under uniformly
comparable measures, and upper bounds in terms of R\'enyi entropy
growth and mass dimension. The concentration estimates and explicit examples show
that local vertex clustering can strictly reduce uniform-measure
energy dimensions without changing the mass dimension.

We have also investigated the role of uniform Ahlfors regularity. 
Under this condition, the optimized \(q\)-energy and the \(q\)-energy of the uniform counting measure admit two-sided estimates in terms of an explicit integral. 
This yields an integral criterion for energy boundedness and determines the critical energy dimensions.

Further questions include identifying broader geometric conditions
under which the uniform and optimized dimensions coincide, and
determining when the entropy upper bounds are sharp. It would also
be useful to compare these energy dimensions with dimensions
defined through graph coverings, and to study their stability
under controlled changes of the graph metric. 

\section*{Acknowledgments}
This work was partially supported by the National Natural Science Foundation of China (Nos. 12301107, 62450131, and 12271038) and the Shandong Province Natural Science Foundation, China (No. ZR202209010046).

\section*{Author Contributions}
Jiantao Liu: Methodology; Formal analysis; Writing -- original draft. 
Cheng Zeng: Conceptualization; Methodology; Funding acquisition; Writing -- original draft.
Yumei Xue: Writing -- review \& editing; Supervision; Funding acquisition. 
Qingcheng Zeng: Writing -- review \& editing; Validation. 
All authors have read and agreed to the published version of the manuscript.

\section*{Conflicts of Interest}
The authors declare no conflicts of interest.

\section*{Use of generative AI} During preparation of this manuscript, the authors used ChatGPT to assist with language revision. Responsibility for the mathematical arguments, citations, and final text rests with the authors.

%% ============================================================================
%%  参考文献
%% ============================================================================

\bibliographystyle{unsrt}
\bibliography{bibliography}

\end{document}